\documentclass[preprint,aps]{revtex4}
\usepackage[dvipsnames]{xcolor}
\usepackage{graphicx} % Required to include images
\usepackage{color} % Required for custom colors
\usepackage{amsmath,amssymb,theorem} % Math packages
\usepackage{epsfig}
\usepackage{hyperref} % Required for links and changing link options
\usepackage[german,english]{babel}
\usepackage{chemfig}
\usepackage{mhchem}
\usepackage{cases}

\begin{document}
\title{Molecular Assembly Lines in Active Droplets}

\author{Tyler Harmon$^{1,2}$* and Frank J\"ulicher$^{1,3,4}$*} \affiliation{{$^1$Max Planck Institute for the Physics of Complex Systems, N\"othnitzerstr. 38, 01187 Dresden,
Germany, \\
$^2$Max Planck Institute of Molecular Cell Biology and Genetics, Pfotenhauerstr. 108, 01307 Dresden,\\
$^3$Center for Systems Biology
Dresden, 01307 Dresden, Germany.\\
$^4$Cluster of Excellence Physics of Life, TU Dresden, 01062
Dresden, Germany.} }

\begin{abstract}

Large protein complexes are assembled from protein subunits to form a specific structure. In our theoretic work, we propose that assembly into the correct structure could be reliably achieved through an assembly line with a specific sequence of assembly steps.  Using droplet interfaces to position compartment boundaries, we show that an assembly line can be self organized by active droplets.  As a consequence, assembly steps can be arranged spatially so that a specific order of assembly is achieved and incorrect assembly is strongly suppressed.
\end{abstract}

\maketitle

\section{Introduction}

Large complexes play important roles in several critical aspects of life. Examples are bacterial flagellar motors, viral capsids, proteasomes, and ribosomes.  Many of such complexes must be assembled into a specific arrangement in order to function.  An important question is how such complexes can be created at high efficiency and fast rate, avoiding the formation of incorrect and incomplete assemblies \cite{Gartner2020,Grant2011,Hagan2011,Chen2008,Hagan2010,Lazaro2016,Reinhardt2014,Jacobs2015,Ke2012,Jacobs2015a,Whitelam2015,Sear2007,Wei2012,Rotskoff2018}.  One strategy of self assembly is complex formation through a thermodynamic process where the assembled state is a free energy minimum \cite{Hagan2011,Grant2011}.  There could exist several accessible minima such that self assembly  would not lead to a unique structure \cite{Gartner2020,Hagan2011,Grant2011,Whitelam2015}.  Furthermore, assembly can often be slowed by trapping in metastable intermediates.  An example of complexes that have been proposed to employ a thermodynamic assembly strategy are viral capsids \cite{Chen2016}.  

Alternative strategies based on non-equilibrium assembly could avoid the problems of thermodynamic self assembly such as kinetic traps and allow for the assembly of structures that cannot be reached by minimizing the free energy \cite{Gartner2020,Whitelam2015,Rotskoff2018}.  This can for example be achieved when components assemble in a specific temporal order.  One of these strategies is separation of assembly timescales.  Here the assembly of the components that need to be added early have high association kinetics while components that need to be added late have slow association kinetics.  This biases the addition of subunits towards a correct temporal ordering.  For this strategy to yield a robust temporal ordering, it requires kinetics that span many orders of magnitude.  Another non-equilibrium strategy uses a carefully timed sequential synthesis to provide subunits when specific assembly phases have completed.  In this process, incorrect binding of subunits is prevented by allowing them to encounter the nascent complex when only the correct binding site is accessible.  This strategy requires specific gene expression programs which limit the rate of complex formation.  It is for example employed in the assembly of the bacterial flagella motor \cite{Alon2006}.

These examples of non-equilibrium assembly lead us to the question of whether complex assembly could emerge as a non-equilibrium pattern in a self-organized process.  More precisely, we ask whether reliable and high-throughput assembly of a complex with a specific structure can be achieved in a steady state where components flow in and complexes flow out.  We show that this is possible in a self-organized assembly line.  The assembly line organizes assembly steps in space, thereby providing a specific temporal order during complex assembly.  Such an assembly line could be organized within a biochemical compartment.

Biochemical compartments in cells either have a membrane or are membraneless.  Membraneless compartments are often liquid like biomolecular condensates that can be described as droplets that phase separate from their surroundings \cite{Li2012,Banani2017,McCall2018,Weber2019}.  A rich diversity of geometries for coexisting droplets have been observed in cells.  One example is a smaller droplet inside a larger droplet \cite{Feric2016,Boisvert2007}.  It remains an open question of what functions such an arrangement of compartments can have.

Here we present the theory of non-equilibrium complex formation by a self organized assembly line.  These emerge as patterns in an active reaction-diffusion system that is confined in droplets. These patterns consist of separated bands of distinct reactions which correspond to different assembly steps.  Chemical patterns are typically viewed through the lens of patterns of concentrations.  Note that the bands discussed here reflect the localization of distinct reactions while the molecular species are not localized.  The spatial arrangement of different assembly steps defines the temporal order in which subunits are added to the complex. This process can occur at steady state with a constant influx of subunits and constant outflux of finished complexes.  This scenario allows a high rate of assembly and ensures the formation of correctly assembled complexes.  

We first develop these ideas using a simplified one dimensional model where the position and width of the reaction bands can be calculated analytically.  We consider a 1D geometry of three distinct volumes separated by two boundaries, see figure 1a.  These volumes could correspond to an arrangement of droplets.  Subunits enter the central stage volume from the left and right volumes.  Assembly of complexes occurs in the stage volume, and is considered irreversible.  We show that if the assembly is fast, association happens at a specific location in the stage volume.  This location depends on the physical properties of the components and their supply at the boundaries. 

We then consider a simplified model in 3-dimensions based on two concentric droplets, see figure 1b.  Here the stage volume is defined by a spherical droplet that coexisting with both an outer liquid phase and an inner droplet.  Components enter the stage via the boundaries from opposite directions.  Such a geometry of droplets inside droplets can be found in cells, for example in the nucleolus \cite{Boisvert2007,Weber2015,Feric2016}.

 \begin{figure}
 	\begin{center}
 		\includegraphics[trim=135 0 135 0, angle=-90,clip,width=\columnwidth]{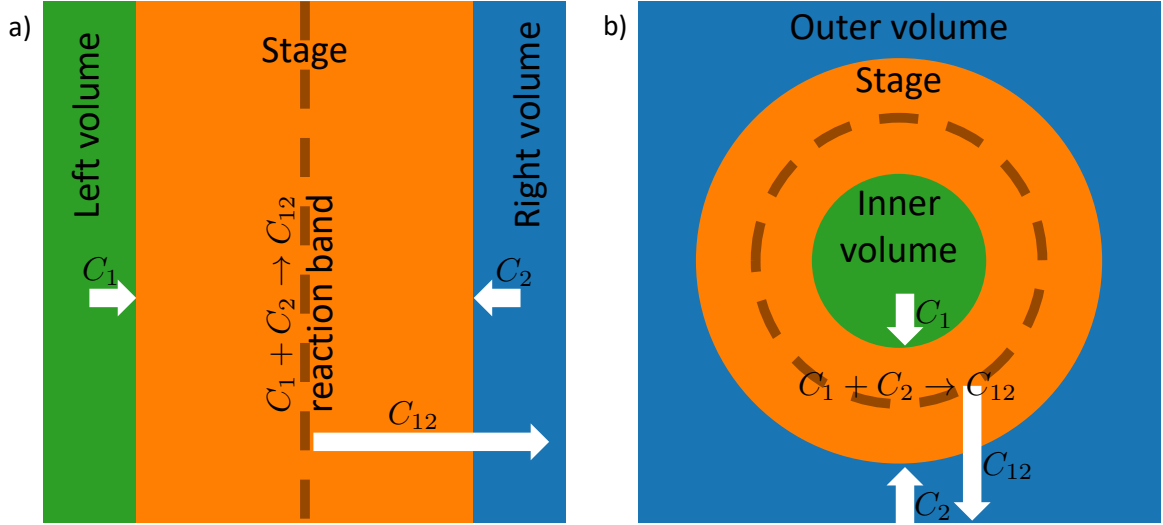}
 	\end{center}
 	\caption{
 		Spatial organization of macromolecular assembly.  a) 1D arrangement of compartments for molecular assembly.  Component $C_1$ is produced in the left volume (green) and component $C_2$ is produced in the right volume (blue).  Assembly of the complex $C_{12}$ occurs in the stage volume (orange).  Assembly can be localized to a band between the left and right volumes (dashed line).  The complex is exported via the right volume.  b) 3D arrangement of compartments for molecular assembly.  Component $C_1$ is produced in the inner volume (green) and component $C_2$ is produced in the outer volume (blue).  Assembly of the complex $C_{12}$ occurs in the stage volume (orange).  Assembly can be localized to a spherical band between the inner and outer volumes (dashed line).  The complex is exported via the outer volume.}
 	\label{fig_1}
 \end{figure}

\section{Localizing assembly steps in one dimension}
First we examine a minimal model in one dimension where molecular assembly steps can be positioned along a line and separated in space.  The region, $[0,L]$, corresponds to the stage where assembly steps occur.  The left boundary connects the stage to the inner volume, and the right boundary connects the stage to the outer volume which serve as reservoirs that provide components.  We consider two components $C_1$ and $C_2$, which enter from the left and right boundaries respectively.  These components bind irreversibly to form a complex,
\begin{equation}
C_1+C_2\rightarrow C_{12} \nonumber\;.
\end{equation}
After the formation of the complex $C_{12}$ it will leave the stage into the outer volume via diffusion.  All molecular species move by diffusion and associate following mass action kinetics.  The system is thus described by the association diffusion equations along the position coordinate x given by 
\begin{subequations}
\begin{align}
\frac{\partial}{\partial t}n_1&=D\frac{\partial^2}{\partial x^2}n_1-kn_1n_2\;,\label{1d_deq1}\\
\frac{\partial}{\partial t}n_2&=D\frac{\partial^2}{\partial x^2}n_2-kn_1n_2 \;.\label{1d_deq2}
\end{align}
\end{subequations}
Here $n_{1}$ and $n_{2}$ are the concentrations of components $C_{1}$ and $C_{2}$ respectively and $D$ and $k$ are the diffusion constant and association rate.  For simplicity we have chosen equal diffusivities for both species.  These equations are complemented by boundary conditions that connect the stage to the inner and outer volumes.  We consider the case where there are no sinks in the inner volume and $C_1$ and $C_2$ are produced only in the inner and outer volumes respectively.  Accordingly, we impose constant influx $J$ of component $C_1$ and no flux of component $C_2$ at $x=0$.  At $x=L$ we impose boundary conditions that depend on the concentrations $n_{1}$ and $n_{2}$ which describe a partitioning of molecules between the stage and the outer volume.  The boundary conditions then read
\begin{subequations}
\begin{align}
	D\partial_x n_1\rvert_{x=0}=&-J\;,\label{1dbc_n1_0}\\
	D\partial_xn_2\rvert_{x=0}=&0\;,\label{1dbc_n2_0}\\
	D\partial_xn_1\rvert_{x=L}=&-\beta_1n_1(x=L)\;,\label{1dbc_n1_L}\\
	D\partial_xn_2\rvert_{x=L}=&\alpha-\beta_2n_2(x=L)\;.\label{1dbc_n2_L}
\end{align}
\end{subequations}
Here $\alpha$ corresponds to the source of $C_2$ and $\beta_1$ and $\beta_2$ correspond to sinks of $C_1$ and $C_2$ respectively.

\begin{figure}[h]
	\begin{center}
		\includegraphics[angle=-90,width=\columnwidth]{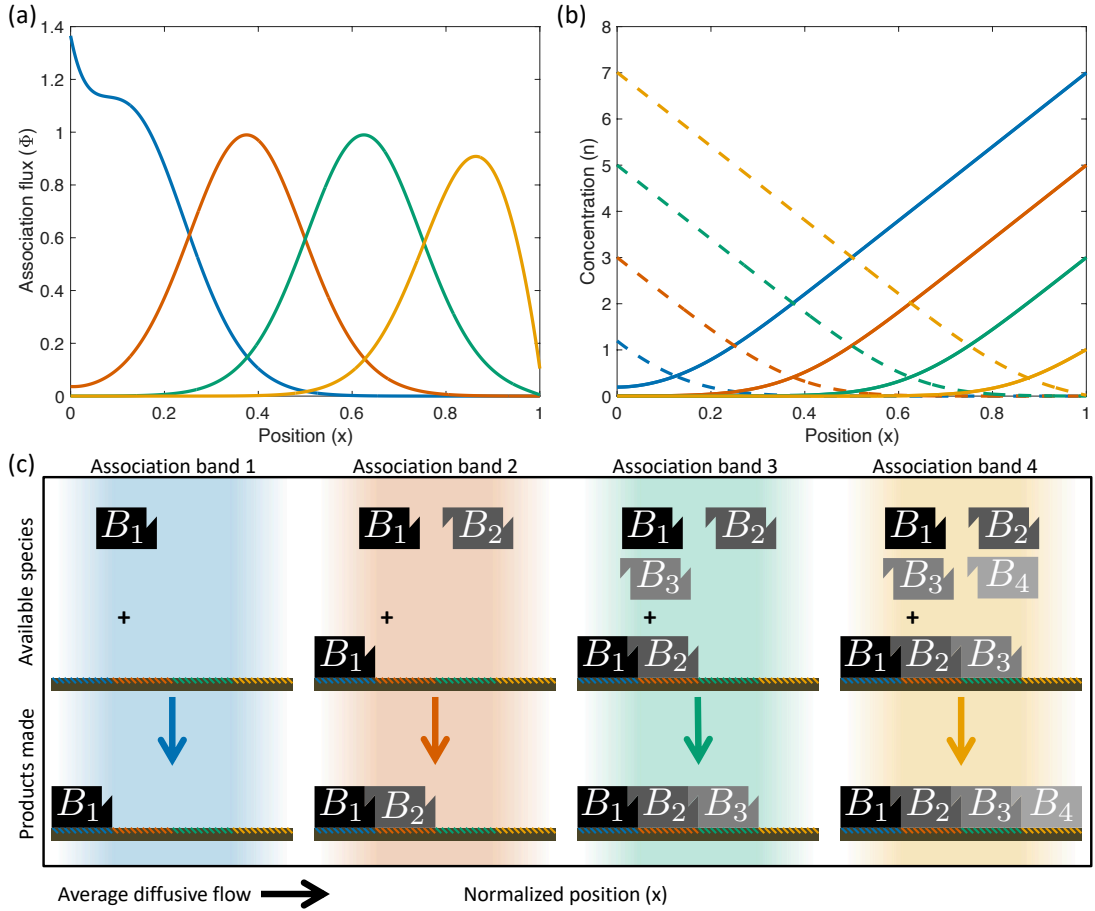}
	\end{center}
	\caption{
		Positioning of assembly steps in narrow regions.  (a) Normalized assembly flux $\Phi$ as a function of normalized position x in the stage region for different values of the normalized influx $\alpha$ of component $C_2$ ($\alpha=70$ (blue), $50$ (orange), $30$ (green), $10$ (yellow)).  (b) Steady state concentrations of components $C_1$(dashed) and $C_2$(solid) for the same values of $\alpha$ as in (a).  Shown are numerical solutions of the minimal model given by equations (2) and (3).  (c) Cartoon schematic of how an assembly line could be organized.  The foundation molecule tends to diffuse to the right progressing through the different association bands.  Progressing to each sequential association bands adds another available brick that can bind to the foundation.  Parameter values are $J=0.31$, $D=0.039$, $k=5.9$ $\beta_1=5$, $\beta_2=10$, and $L=1$.}
	\label{fig_2}
\end{figure}

Of particular interest to us is the steady state solution of equations (1) with boundary conditions (2), and the profile of the association flux $\Phi\equiv kn_1n_2$ as a function of $x$.  In figure \ref{fig_2}a we show examples of stationary profiles of the association flux that reveal the localization of the association to different positions depending on the magnitude of $\alpha$, the source of $C_2$.  Interestingly, by changing the magnitude of of $\alpha$, the location $x_M$ of the peak of the association flux is changed, but the shape remains essentially unchanged.  Figure \ref{fig_2}b shows the corresponding steady state concentration profiles for component $C_1$ and $C_2$ as dashed and solid lines, respectively.  

Away from the region of high association flux, the concentration profiles are linear, corresponding to a constant flux.  These fluxes for the two components are equal and opposite because assembly consumes an equal number of both components.  As a result equation (1) has symmetric solutions with respect to $x_M$ if the boundaries are far from the assembly volume.  In this limiting case, assembly is fully efficient within $x=[0,L]$, and we can calculate the exact position $x_M$ and the variance $\sigma^2$ of the association flux profile in a Gaussian approximation, see appendix,
\begin{equation}
x_M=L-D\frac{\alpha-J}{J\beta_2}\;,
\end{equation}
\begin{equation}
\sigma\approx\sqrt{\frac{2}{\pi}}(1+\frac{\pi}{2}+\sqrt{1+\pi})^{1/3}\bigg(\frac{D^2}{kJ}\bigg)^{1/3}\;.
\end{equation}
The limit where boundaries are far from the assembly volumes corresponds to $\sigma\ll x_m$ and $\sigma\ll L-x_m$.

This result shows that one can spatially localize an assembly step in a region of width $\sigma$ centered at position $x_M$. This localization of assembly steps can be used to generate an assembly line.  In this assembly line, the complex grows as it diffuses towards one boundary which we choose to be the one on the right side.  Complex assembly is initiated by a molecule that enters the stage from the left.  It first encounters a single component from the right to which it can bind.  As the assembling complex diffuses toward the right, it will encounter sequentially additional components that arrive from the right boundary.  These components can thus be incorporated sequentially to ensure correct assembly.  This is schematically represented in figure 2c.  We call the molecule entering from the left and initiating complex formation the foundation, denoted $F$, and the components entering from the right the assembly bricks, denoted $B_i$.  Assembly regions can be arranged in space such that they are separated by more than a distance $\sigma$. This implies that the addition of one type of component is completed before the complex encounters the next component. The spatial order of assembly regions specifies the dominant temporal order of assembly steps. This assembly line can operate at steady state with all assembly steps being conducted in parallel.

\section{Assembly lines in spherical geometry}

We now discuss an example of how such an assembly line could be realized in three dimensions using concentric spherical droplets where a small droplet is located in the center of a larger droplet, see figure 1b.  The foundational component, denoted $F_0$ is produced inside the small central droplet which defines the inner volume.  In our example we consider four assembly bricks, $B_i$, $i=1...4$, which are provided in the outer volume that surrounds both droplets.  The stage volume is the space in the larger droplet between the inner volume and outer volume.

The foundation and bricks enter the stage from opposite sides via diffusion.  For simplicity we consider a simple model where the assembly bricks must bind to the foundation in a specific sequential order for correct assembly.  We call the intermediate assemblies assembled in the correct sequential order the on-pathway intermediates, denoted by $F_i$ where $i$ is the index of the last brick that was incorporated into the complex.  In this logic, $F_4$ is the correctly assembled final complex.  If binding occurs in a different order then complex formation is unsuccessful and we call these configurations off-pathway.  Figure 3a shows a schematic of all constructs and associations, both on and off-pathway.  We define our model using concentration fields for all components and assembly intermediates.  Inside the stage these concentrations are described by association-diffusion equations that capture the diffusion of components and their association steps,
\begin{align}
\frac{\partial}{\partial t}&n_{F,i}=D_{F,i} \nabla^2n_{F,i}-kn_{F,i}n_{B,i+1}
-kn_{F,i}\Theta(2-i)\sum_{k=i+2}^{4} n_{B,k}+kn_{F,i-1}n_{B,i}\;,\label{3d_deq1}\\
\frac{\partial}{\partial t}&n_{B,i}=D_{B,i}\nabla^2n_{B,i}-kn_{F,i-1}n_{B,i}
-kn_{B,i}\Theta(i-2)\sum_{k=0}^{i-2} n_{F,k}\;.\nonumber
\end{align}
Here $n_{B,i}$ is the concentration of brick $i$, $n_{F,i}$ is the concentration of the free foundation ($i=0$), the on-pathway intermediates ($i=1,2,3$), and product ($i=4$). Furthermore $D_{F0}$, $D_{Fi}$ and $D_{Bi}$ are the diffusion constants for foundation, complexes and bricks respectively, and $k$ is the rate constant for on and off pathway associations.  The Laplace operator is denoted $\nabla^2$ and the Heaviside function is denoted $\Theta$ with $\Theta(i)=1$ if $i\geq 0$ and $\Theta(i)=0$ otherwise.  The first term of the right hand sides in equations 5 describe diffusion, the second terms describe the component loss due to on-pathway assembly, the third terms describe loss via off-pathway associations that does not lead to correct assembly in our model, and the fourth term is a source due to on-pathway assembly.  These equations are supplemented by boundary conditions at radii $r_1$ and $r_2$,
\begin{subequations}
\begin{align}
\frac{dn_{F,0}}{dr}\biggr\rvert_{r_1}=&-\frac{J}{D_F4\pi r_1^2}\;,\label{3d_bc1}\\
\frac{dn_{F,i}}{dr}\biggr\rvert_{r_2}=&-\frac{\beta_F n_{F,i}(r_2)}{D_F4\pi r_2^2}\;,\label{3d_bc2}\\
\frac{dn_{B,i}}{dr}\biggr\rvert_{r_2}=&\frac{\alpha_{B,i}-\beta_{B,i} n_{B,i}(r_2)}{D_{B,i}4\pi r_2^2}\;,\label{3d_bc3}
\end{align}
\end{subequations}
where $J$ is the number of foundation molecules produced per unit time in the inner volume, $\alpha_{Bi}$ is the influx of brick $B_i$ from the outer volume, $\beta_F$ and $\beta_{B,i}$ describes the concentration dependence of the outflux of the components into the outer volume due to partitioning.  Furthermore, we have imposed no flux boundary conditions at $r_1$ for intermediates $F_i$ where $i\geq1$ and the bricks $B_i$.  These expressions can be derived for a droplet in a steady state environment. 

In spherical coordinates we define the total association flux as
\begin{equation}
\Phi_i=4\pi r^2kn_{F,i-1}n_{B,i}\;.\label{3d_phi_def}
\end{equation}  
In steady state the total association flux exhibits a maximum at position
\begin{equation}
r_{M,i}\approx r_2-\frac{4\pi r_2^2D_{B,i}}{\beta_{B,i}}\bigg(\frac{\alpha_{B,i}-J}{J}\bigg)\;,\label{r_Mi}
\end{equation}
and a width given by 
\begin{equation}
\sigma_i\approx 1.33\bigg(\frac{4\pi r_{M,i}^2D_FD_{B,i}}{kJ}\bigg)^{1/3}\;,\label{sigma_i}
\end{equation}
see appendix.  Figure 3b shows the fraction of complete, incomplete, and incorrect assemblies leaving the stage at steady state as a function of association rate.  The fraction of complete assemblies defines the efficiency of the system. Figure 3c and d show the radial profiles of the total association fluxes for two different association rates.  The association rate primarily changes the width of the peaks of the total association flux but not the position, see equations \eqref{r_Mi} and \eqref{sigma_i}.  For more sharply peaked total association fluxes, the efficiency of assembly is higher because different assembly steps are better separated in space.  Accordingly, assembly efficiency increases as the association rate increases and off-pathway associations are highly suppressed, see figure 3b.  The general requirement for assembly lines to improve assembly fidelity is that the spacing between association bands is larger than the width of the bands.  Note that in spherical geometry the assembly steps occur in concentric shells around the inner volume.

\begin{figure}
	\begin{center}
		\includegraphics[angle=-90,width=15cm]{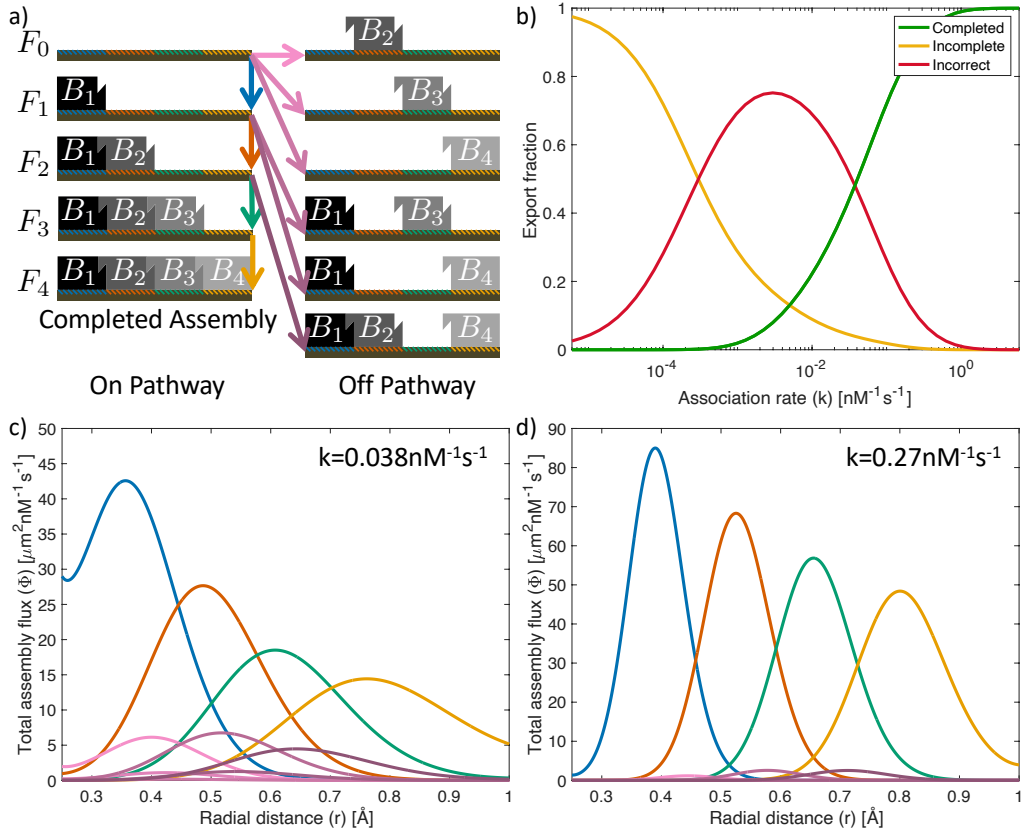}
	\end{center}
	\caption{
		 Self-organized molecular assembly lines.  (a) Scheme of the on pathway assembly steps (blue, red, green, and yellow arrows) and off-pathway dead ends (pink arrows).  Here $F_0$ represents the foundation and $B_1$, $B_2$, $B_3$, $B_4$ represent the bricks.  The on-pathway intermediates are denoted $F_1$, $F_2$, $F_3$, and $F_4$ is the correctly assembled product.  (b) Fraction of completed (green), incomplete (yellow), and incorrect (red) assemblies that leave the stage in steady state as a function of the association rate $k$.  (c) Radial profiles of total assembly fluxes $\Phi=4\pi r^2kn_{F,i}n_{B,j}$ at steady state for the four on-pathway associations (blue, red, green, yellow) and the off-pathway associations (pink) for association rate $k=0.038\;\rm{nM}^{-1}\rm{s}^{-1}$.  Here $n_{F,i}$ and $n_{B,j}$ are concentrations of associating components.  (d) same as (c) but for $k=0.27\;\rm{nM}^{-1}\rm{s}^{-1}$.  The curves shown in (c) and (d) correspond to the assembly steps shown in (a) in the same color.  Shown in (b-d) are numerical solutions for spherical geometry and parameter values $J=10\;\rm{\mu m}^3\rm{nMs}^{-1}$, $\beta_F=1.26\;\rm{\mu m}^{3}\rm{s}^{-1}$, $\beta_B=1.26\;\rm{\mu m}^{3}\rm{s}^{-1}$, $r_1=0.25\;\rm{\mu m}$, $r_2=1.0\;\rm{\mu m}$, $D_F=0.01\;\rm{\mu m}^2\rm{s}^{-1}$, $D_{B,i}=0.01\;\rm{\mu m}^2\rm{s}^{-1}$, and $\alpha_{B,i}=(276,163,100,57)\;\rm{\mu m}^3\rm{nMs}^{-1}$ for $i=(1,2,3,4)$.}
	\label{fig_3}
\end{figure}

\section{Discussion}

Cells need to assemble molecular complexes in a correct and efficient manner and avoid off-pathway dead ends.  We have shown that molecular assembly lines can be self-organized in cellular compartments such as phase separated condensates.  Such an assembly line requires a spatial organization of association steps which distinguishes it from well mixed association scenarios.  This spatial organization can lead to a high-throughput assembly by operating at steady state.  Additionally, high-fidelity of assembly can be achieved by controlling the temporal order in which subunits are added.  Control over temporal order is possible by the spatial separation of the assembly bands.  The spatial arrangement of these bands could be controlled by concentrations of components at their sources, which we note can be a distance from the assembly line.  The scenario proposed here has four hallmarks that could be seen in experiments: i) association but not concentration is confined to distinct spatial bands, ii) unbound components can be found close to their source but not past the position where they are added to the complex, iii) the average complex size increases towards the outer boundary, and iv) changing the concentration ratio of components yields incomplete or incorrectly assembled complexes because assembly bands can become rearranged.

Self organized molecular assembly lines can be compared to assembly under well mixed conditions where concentrations are constant in space.  Using the parameter values given in the caption for figure 3b, but considering assembly in well mixed conditions, we find an export fraction of correct assemblies $E=\prod_i^4 kn_{\infty,i}/\sum_{j=i}^4kn_{\infty,j}\simeq 15\%$.  This is much smaller than values approaching $100\%$ which can be obtained in the self-organized assembly line for the same parameter values, see figure 3b. This reduction in export fraction of correct assemblies occurs because intermediates are exposed to all the binding bricks and incorrect binding steps therefore become likely.  This contrasts with an efficient assembly line where binding bricks are added sequentially in the correct order.  In well mixed conditions, high fidelity can be achieved if time scales are separated such that different assembly steps occur at different rates.  In particular, components that need to bind early should have the fastest association kinetics.  However, in order to achieve an export fraction $E$ of correctly assembled complexes, this strategy requires association rates that span at least a factor of $(E^{-1/(N-1)}-1)^{N-1}$ where $N$ is the number of assembly bricks in the complex, see appendix.  For $80\%$ efficiency of assembly with four bricks, matching our assembly line example, the association rates would need to span a factor of 2200.  Such a large range of rates suggests that some steps must be slow and thus rate limiting for assembly.  This is in contrast to the assembly line proposed here where the example in figure 3 uses parameters that span a factor of 5.  Indeed, a large span of assembly rates manifests as a equally large range of concentration levels at steady state.  Therefore, the well mixed scenario requires much larger amounts of unfinished complexes to achieve the same export rate.  This consumes more resources and materials and is therefore less efficient.

For simplicity we considered simplified geometries in one and three dimensions, using a spherical geometry motivated by liquid-like condensates.  However, the strategy discussed here is more general and can also work in other geometries.  As illustrated in the one dimensional case, the assembly steps are positioned at specific distances from the outer surface.  Therefore, our work suggests that in non-spherical geometries, assembly steps will be positioned to non-intersecting manifolds inside the outer surface.   We expect this mechanism to robustly self organize assembly lines in three dimensions in a broader collection of geometries, including multiple spherical inner volumes.

The assembly line described here can control the temporal order of assembly steps.  In principle, complexes can also form without a fixed temporal order.  For example, if many identical components, such as a viral capsid, then the time order is irrelevant.  On the other extreme, if all components are different then temporal order can be helpful.  In this case it is likely that several local minima exist which correspond to different assembled structures.  In principle a unique structure emerges as the lowest energy configuration but this may take a long time to reach.  However, if the temporal order is controlled, a specific local minima could be consistently reached.  Therefore time ordered assembly provides a strategy to reliably assemble many components into a specific structure.  For simplicity we have considered irreversible assembly steps.  In practice molecular binding events are not fully irreversible and the limit of irreversible steps corresponds to a situation where the time scale for an unbinding event is slower than the time between subsequent assembly steps.  In this case unbinding events become rare enough to become negligible.

An important example of molecular assembly are RNA-protein complexes.  The components of such complexes can fit well into our simplified assembly scheme, shown in figure 3a.  Here bricks (proteins) are added to a foundational element (RNA) and the bricks and foundation are supplied from different regions (translation in the cytoplasm and transcription in the nucleus respectively). 

The most prominent example of an RNA-protein complex is the ribosome which mainly consists of large RNA and many different small proteins \cite{Tschochner2003,Davis2017,Klinge2019}.  Ribosomes are assembled in the nucleolus, a liquid-like droplet-inside-droplet compartment in the cell nucleus.  Ribosomal RNA is produced in the inner compartment of the nucleolus called the Dense Fibrillar Component, corresponding to the inner volume.  Ribosomal proteins are provided in the nucleoplasm, corresponding to the outer volume.  The assembly occurs in the Granular Component of the nucleolus corresponding to the stage volume.  Given this arrangement, ribosome assembly is an ideal candidate for the molecular assembly line proposed here.  It will be interesting to explore whether signatures of an assembly line can be found in ribosome assembly.  An interesting question is whether assembly bands could be directly observed experimentally.  This poses a challenge because labeling components alone may not reveal the spatial organization of an assembly line.  This is because florescent labels would be attached to a component as well as to the assembled complex.  Even if the assembly is organized into distinct bands, the florescent signal would not show this structure.  Techniques such as FRET could differentiate between bound and unbound components and thus are promising to reveal spatial patterns of assembly processes.

In addition to the nucleolus, other liquid like compartments could play a role to organize the assembly of macromolecular complexes.  An example is p-granules in \textit{C. elegans} which can be located on the nuclear membrane and cover nuclear pores \cite{Updike2011}.  This setting has the important feature that p-granules can communicate with two different compartments at the same time, the cytoplasm and nucleoplasm, allowing opposing fluxes from both compartments.  This suggests that p-granules are also naturally suited for reliable molecular processing using assembly lines of the type proposed here.  More generally, we expect that liquid like compartments are used to self organize biochemical processes in space.  Our work therefor provides new insights into the possible role of liquid like condensates for biological processes.

\section{acknowledgments}

We thank Rabea Seyboldt, Anthony A. Hyman, Richard W. Kriwacki, and Diana Mitrea for useful discussions.

\end{document}

\title{Molecular Assembly Lines in Active Droplets}

\author{Tyler Harmon$^{1,2}$, Frank J\"ulicher$^{1,3,4}$} \affiliation{{$^1$Max Planck Institute for the Physics of Complex Systems, N\"othnitzerstr. 38, 01187 Dresden,
Germany, \\
$^2$Max Planck Institute of Molecular Cell Biology and Genetics, Pfotenhauerstr. 108, 01307 Dresden,\\
$^3$Center for Systems Biology
Dresden, 01307 Dresden, Germany.\\
$^4$Cluster of Excellence Physics of Life, TU Dresden, 01062
Dresden, Germany.} }

%\centerline{date}
%\centerline{Tyler Harmon}

\setcounter{equation}{9}

\maketitle

\section{Approximate solution in One Dimension}

From integrating $n_1''=n_2''$ we obtain 
\begin{equation}
n_1+a(x-x_M)=n_2-a(x-x_M)\;.
\end{equation}
We define these as $\theta$ because when substituted into equation (1a),
\begin{subequations}
	\begin{align}
	\theta=&n_1+a(x-x_M)\;,\label{1d_theta_def}\\
	\theta''=&\theta^2-a^2(x-x_M)^2\label{1d_theta_deq}\;,
	\end{align}
\end{subequations}
highlights the symmetry of the differential equations around $x=x_M$.  Using \eqref{1d_theta_def} and considering the full consumption limit where $n_1(L)\rightarrow 0$, we can rewrite our boundary conditions in equation (2) in the form,
\begin{subequations}
\begin{align}
	\partial_x \theta\rvert_{x=0}= &-\frac{J}{D}+a\;,\label{1dbc_n1_0}\\
	\partial_x\theta\rvert_{x=0}= &-a\;,\label{1dbc_n2_0}\\
	\partial_x\theta\rvert_{x=L}= &a\;,\label{1dbc_n1_L}\\
	\partial_x\theta\rvert_{x=L}= &\frac{\alpha}{D}-\frac{2a\beta_2}{D}(L-x_M)-a\;.\label{1dbc_n2_L}
\end{align}
\end{subequations}
From equation \eqref{1dbc_n2_0} and \eqref{1dbc_n1_L} it is clear that the boundary conditions for $\theta$ are also symmetric.  With symmetric differential equations and symmetric boundary conditions, $\theta$ must be symmetric.  Thus if there is a single maximum in the association flux $\Phi=D\theta''$ then $x_M$ must be the location of this maximum.  Equating the first two boundary conditions gives us
\begin{equation}
a=\frac{J}{2D}\;.\label{1d_a}\;,
\end{equation}
and equating the last two boundary conditions gives us
\begin{equation}
x_M=L-\frac{D}{\beta_2}\frac{\alpha-J}{J}\;.
\end{equation}
For values of $x_M$ between $0$ and $L$, $x_M$ is the position of the maximum association flux.  This corresponds to the association being targeted to a specific region inside the stage.  We can also estimate the width of the association volume, or how well the association is restricted to that position by approximating $\Phi$ as a normal distribution with a width of $\sigma$,
\begin{align}
\log(\Phi)=&\log(\frac{J}{\sigma\sqrt{2\pi}})-\frac{(x-x_M)^2}{2\sigma^2}\;,\\
\log(\Phi)=&\log(k\theta^2-ka^2(x-x_M)^2)\;.
\end{align}
Taylor expanding $\log(\Phi)$ around $x_M$ gives us
\begin{equation}
\log(\Phi)=\log(k\theta^2(x_M))+\frac{\theta(x_0)\theta''(x_M)-a^2}{\theta^2(x_M)}(x-x_0)^2+O((x-x_M)^4)\;.
\end{equation}
Equating terms and substituting in equation \eqref{1d_theta_deq} gives us
\begin{align}
k\theta^2(x_M)=&\frac{J}{\sigma\sqrt{2\pi}}\;,\\
-\frac{1}{\sigma^2}=&2\theta(x_M)-\frac{J^2}{2\theta^2(x_M)}\;.
\end{align}
Solving for $\sigma$ gives us the width of our association volume,
\begin{equation}
\sigma=\sqrt{\frac{2}{\pi}}\bigg(1+\frac{\pi}{2}+\sqrt{1+\pi}\bigg)^{1/3}\bigg(\frac{D^2}{kJ}\bigg)^{1/3}\approx 1.33\bigg(\frac{D^2}{kJ}\bigg)^{1/3}\;.
\end{equation}

\section{Approximate solution in spherical coordinates}

We begin by making several approximations consistent with our minimal model.  First, that the associations happen away from the interfaces.  Because the associations happen at a rate proportional to both concentrations, this means that the brick concentration must be zero at the inner interface and the foundation concentration must be zero at the outer interface.  Additionally, we assume that the on-pathway associations are all well separated in space.  This means we can neglect all off pathway associations and can solve each brick concentration independently from the rest.  This reduces the steady state problem to
\begin{align}
D_{F}\nabla^2n_{F}=&kn_{F}n_{B,i}\;,\label{3d_a_deq1}\\
D_{B,i}\nabla^2n_{B,i}=&kn_{F}n_{B,i}\;,\nonumber
\end{align}
with boundary conditions
\begin{subequations}
	\begin{align}
	\frac{dn_{F}}{dr}\biggr\rvert_{r_1}=&-\frac{J}{D_F4\pi r_1^2}\;,\label{3d_a_bc1}\\
	\frac{dn_{B,i}}{dr}\biggr\rvert_{r_1}=&0\;,\label{3d_a_bc2}\\
	\frac{dn_{F,i}}{dr}\biggr\rvert_{r_2}=&-\frac{\beta_{F} n_{F}(r_2)}{D_F4\pi r_2^2}\;,\label{3d_a_bc3}\\
	\frac{dn_{B,i}}{dr}\biggr\rvert_{r_2}=&\frac{\alpha_{B,i}-\beta_{B,i} n_{B,i}(r_2)}{D_{B,i}4\pi r_2^2}\;.\label{3d_a_bc4}
	\end{align}
\end{subequations}
From integrating $D_F\nabla^2n_F=D_{B,i}\nabla^2n_{B,i}$ from equation \eqref{3d_a_deq1} we have
\begin{equation}
rD_Fn_F+a_i(r-r_{M,i})=rD_{B,i}n_{B,i}-a_i(r-r_{M,i})\;.
\end{equation}
We define this as $\theta_i$ because when substituted into equation \eqref{3d_a_deq1},
\begin{subequations}
	\begin{align}
	\theta_i=&rD_Fn_F+a_i(r-r_{M,i})\;,\label{3d_a_thetadef}\\ %3d_a_thetadef
	\theta_i''=&\frac{k}{rD_FD_{B,i}}(\theta_i^2-a_i^2(r-r_{M,i})^2)\;,\label{3d_a_thetapp}
	\end{align}
\end{subequations}
highlights that as long as the association volume is small compared to $r_{M,i}$ then $\theta_i$ must be symmetric around $r=r_{M,i}$ just like the one dimensional case.  Thus if there is a single maximum in the association flux $\Phi_i=4\pi r\theta_i''$ then $r_M$ must be the location of this maximum. Substituting equation \eqref{3d_a_thetadef} into our boundary conditions at $r=r_1$, equations \eqref{3d_a_bc1} and \eqref{3d_a_bc2}, gives us
\begin{equation}
a_i=\frac{J}{8\pi r_{M,i}}\label{3d_a}
\end{equation}
Considering the full consumption limit where $n_{F,i}(r_2)\rightarrow 0$, substituting equation \eqref{3d_a_thetadef} and \eqref{3d_a} into the boundary conditions at $r=r_2$, equations \eqref{3d_a_bc3} and \eqref{3d_a_bc4}, gives us the position of the maximum association,
\begin{equation}
\frac{1}{r_{M,i}}=\frac{1}{r_2}+\frac{4\pi D_{B,i}(\alpha_{B,i}-J)}{\beta_{B,i}J}\;.
\end{equation}
For values of $r_{M,i}$ between $r_1$ and $r_2$, $r_{M,i}$ is the position of the maximum association flux.  This corresponds to the association being targeted to a specific region inside the stage.  We can also estimate the width of the association volume, or how well the association is restricted to that position by approximating $\Phi_i$ as a normal distribution with a width of $\sigma_i$,
\begin{align}
\log(\Phi_i)=&\log(\frac{J}{\sigma_i\sqrt{2\pi}})-\frac{(r-r_{M,i})^2}{2\sigma_i^2}\;,\\
\log(\Phi_i)=&\log(\frac{4\pi k}{D_FD_{B,i}}(\theta_i^2-a_i^2(r-r_{M,i})^2))
\end{align}
Taylor expanding $\log(\Phi_i)$ around $r_{M,i}$ gives us
\begin{equation}
\log(\Phi_i)\approx\log(\frac{4\pi k}{D_FD_{B,i}}\theta_i^2(r_{M,i}))+\frac{\theta_i(r_{M,i})\theta''(r_{M,i})-a_i^2}{\theta^2(r_{M,i})}(r-r_{M,i})^2
\end{equation}
Equating terms and substituting in equations \eqref{3d_a_thetapp} and \eqref{3d_a} allows us to solve
\begin{equation}
\sigma_i\approx \sqrt{\frac{2}{\pi}}(1+\frac{\pi}{2}+\sqrt{1+\pi})^{1/3}\bigg(\frac{4\pi r_{M,i}^2D_FD_{B,i}}{kJ}\bigg)^{1/3}
\end{equation}

\section{Efficiency of separation of timescales}

We consider a system with $N+2$ components that are assembled into a complex.  We impose the rule that a component can be added to a complex as long its numeric index is higher than the largest component currently added.  In this way, the only complexes with $N+2$ components are ones that were assembled in numeric order.  We use the denote the first component in the unbound state as the $i=0$ complex.  The rate of adding the $i$'th component is $k_i$.  The probability that a complex assembled up to $i-1$ will add the $i$'th component is given by
\begin{equation}
P_{i-1\rightarrow i}=\frac{k_i}{\sum_{j=i}^{N+1}k_j}\;.
\end{equation}
The efficiency of assembly is the probability that every step was assembled in the right order is thus given by the product of each step,
\begin{equation}
E=\prod_{i=1}^N\frac{k_i}{\sum_{j=i}^{N+1}k_j}\;.
\end{equation}
If the reaction rates are evenly spaced on a logarithmic scale then we can express $k_i=\beta\alpha^i$.  This gives a total probability as
\begin{equation}
E=\prod_{i=1}^N\frac{\alpha^i}{\sum_{j=i}^{N+1}\alpha^j}\;.
\end{equation}
If $\alpha\ll 1$ then we can truncate the summation after the second term yielding
\begin{equation}
E=\frac{1}{(1+\alpha)^{N}}\;.
\end{equation}
Thus in order to achieve a given efficiency we need to have a separation between steps
\begin{equation}
\alpha=E^{-1/N}-1\;,
\end{equation}
and a total separation of $\alpha^{N}$.